\documentclass[a4paper,11pt]{article}

\usepackage{contribution}

\newcommand{\weblink}[2][]{%
    \ifthenelse{\equal{#1}{}}%
    {\textnormal{\url{#2}}}%
    {\textnormal{\href{#2}{#1}}}%
}

\newcommand{\acknowledgements}[1]{%
  \bigskip\bigskip
  \textsf{\textbf{\Large Acknowledgements}} \\[2ex]
  {#1}
  \bigskip
}

\def\beq{\begin{equation}}
\def\eeq#1{\label{#1}\end{equation}}
\def\eeqn{\end{equation}}

\def\beqa{\begin{eqnarray}}
\def\eeqa#1{\label{#1}\end{eqnarray}}
\def\eeqan{\end{eqnarray}}

\let\bar=\overbar

\def\Dslash{\not{\hbox{\kern-4pt $D$}}}
\def\dslash{\not{\hbox{\kern-2pt $\del$}}}

\def\msb{{\bar{\ssstyle M \kern -1pt S}}}

\newcommand{\contribution}[7][]{%
  \clearpage
  \thispagestyle{plain}
  \ifthenelse{\equal{#1}{}}
  {\hypersetup{pdftitle={#2}}}
  {\hypersetup{pdftitle={#1}}}
  \hypersetup{pdfauthor={{#3} {#4}}}
  {\centering\normalfont\LARGE\bfseries\sffamily #2 \par\nobreak}
  \lhead{}
  \chead{%
    \textit{\footnotesize XIV International Conference on Hadron Spectroscopy
      (\weblink[\textit{hadron2011}]{http://www.hadron2011.de}), 13-17 June 2011, Munich, Germany}%
  }
  \rhead{}
  \bigskip
  \begin{center}
    {#3} {#4}\ifthenelse{\equal{#6}{}}{}{\footnote{\weblink[#6]{mailto:#6}}}
    \ifthenelse{\equal{#7}{}}{}{#7} \\
    \textit{#5}
  \end{center}
  \bigskip
}

\renewcommand{\abstract}[1]{%
  \begin{center}
    \begin{minipage}{0.85\textwidth}
      \begin{footnotesize}
        #1
      \end{footnotesize}
    \end{minipage}
  \end{center}
  \bigskip
}

\begin{document}

%
%
%
%
%
{  

\makeatletter
\@ifundefined{c@affiliation}%
{\newcounter{affiliation}}{}%
\makeatother
\newcommand{\affiliation}[2][]{\setcounter{affiliation}{#2}%
  \ensuremath{{^{\alph{affiliation}}}\text{#1}}}
  
%

\contribution[Mass dependence of $\pi\pi$ phase shifts: ChPT vs. Lattice]
{Pion mass dependence of $\pi\pi$ phase shifts within standard and unitarized ChPT versus Lattice results}
{Jenifer}{Nebreda}  
{Departamento de F\'isica Te\'orica II. Universidad Complutense de Madrid, 28040, Madrid, Spain
}
{}
{\!\!, Jos\'e Ram\'on Pel\'aez, and Guillermo R\'ios}
%

\abstract{%
  We report on our recent results in the study of the chiral extrapolation of the phase-shifts in elastic pion-pion scattering, using both standard and unitarized ChPT to one and two loops. In the standard ChPT approach, limited to low momenta, we study the S, P and D waves. Unitarization extends the analysis to energies of around 1 GeV, being compatible with standard ChPT at low energies for the S and P waves. We then compare with lattice results and find a good agreement of standard ChPT below 200 MeV for the $I$=2, $J$=0 and $I$=1, $J$=1 channels and up to 500 MeV for the $I$=2, $J$=2 channel. Unitarized ChPT improves the agreement in the scalar and vector channels at higher energies. We have also performed a Montecarlo analysis to provide an estimation of the uncertainties. }
%

\section{Introduction}

Since the low-energy regime is beyond the reach of perturbative QCD, one may rely on lattice techniques in order to describe the hadronic processes in terms of quarks and gluons. However, Lattice QCD presents complications such as the implementation of chiral symmetry, the small physical values of the light quarks and the existence of quarkline disconnected diagrams. Thus, few lattice results on phase shifts are available, and those existing correspond to large pion masses. Fortunately, we can make use of Chiral Perturbation Theory (ChPT), which provides the quark mass dependence of the meson-meson scattering amplitudes in the low energy region, to compare its results to those of lattice studies by increasing the mass of the quarks up to the applicability limits of the theory.

Although in the conference we also commented on the resonance pole dependence on quark masses, here we only refer to the original reference~\cite{Nebreda:2010wv} for such matters and prefer to concentrate on our recent study~\cite{Nebreda:2011di} of the phase shift dependence on the averaged $u$ and $d$ quark mass, $\hat{m}$, or, equivalently, on the pion mass, $M_\pi$. In particular, we include here novel calculations at $M_\pi=266$ MeV in order to compare with very recent lattice results \cite{Lang:2011mn} on the $I$=1, $J$=1 phase shift, since this issue triggered some interesting discussions at this conference.
 
First, we will describe the standard one and two-loop SU(2) ChPT\cite{Bijnens:1995yn}. This approach has the advantage of being completely model independent but it is limited to the low energy region. This is the reason why we next extend our study to unitarized ChPT,  
using the well-known one-loop elastic Inverse Amplitude Method (IAM), which allows us to calculate the phase shift quark mass dependence up to higher values of $M_\pi$.

\section{Quark mass dependence of $\pi\pi$ phase shifts}

\begin{figure} 
\begin{center}
\begin{tabular}{c}
  \includegraphics[scale=1.2]{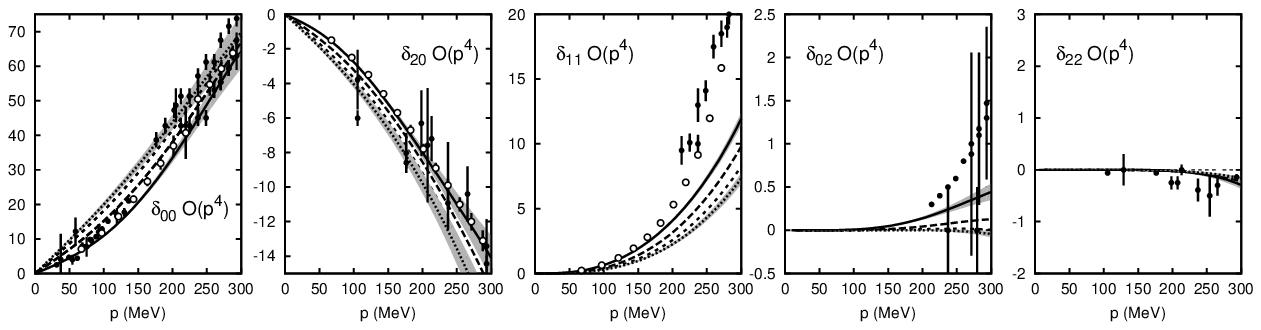}\\
  \includegraphics[scale=1.2]{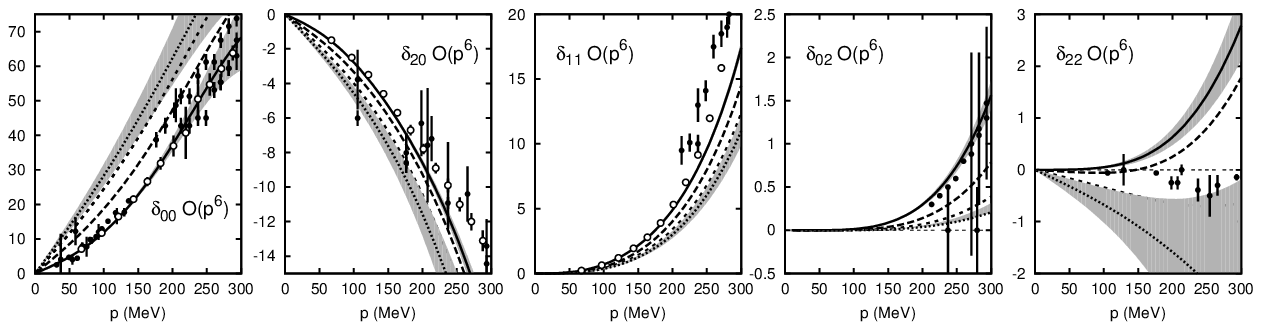}
  \end{tabular}
    \caption{$\pi \pi$ phase shifts from standard ChPT up to one loop (first row) and to two loops (second row). Different lines stand for different pion masses: continuous, long dashed, short dashed and dotted for $M_\pi=139.57,\, 230,\, 300$ and 350 MeV respectively. We only show error bands for the lightest and heaviest masses. Experimental data (circles) come from~\cite{experimentaldata} (black circles) and the precise model independent dispersive data analysis from \cite{GarciaMartin:2011cn} (white circles).
    }
   \label{fig:NUphaseshifts}
  \end{center}
\end{figure}

\paragraph{Standard ChPT} We study the dependence of the $\pi \pi$  phase shifts on the pion mass $M_\pi$ using the SU(2) scattering amplitudes in \cite{Bijnens:1995yn} and the LECs in \cite{Colangelo:2001df}, except for $l_3^r$ that is taken from \cite{Colangelo:2010et}. In Fig.~\ref{fig:NUphaseshifts} we show phases on different channels for different pion masses, to one and two loops (first and second row respectively). They are plotted as a function of the center of mass momentum -and not of the energy- in order to subtract the trivial effect of the threshold. A Montecarlo gaussian sampling based on the errors of the LECs has been used to calculate the error bands. We find that the dependence of the phase shifts on $M_\pi$ is very soft at one loop and somehow stronger at two loops, specially for the $I$=2, $J$=2 channel.

\paragraph{Unitarized ChPT} We use now the elastic IAM to unitarize our amplitudes, so that the applicability limit is extended to the elastic resonance region. In Fig.~\ref{fig:Uphaseshifts} we show $\pi \pi$ phase shifts (note that the one and two-loop IAM cannot be used for the D-waves) for different pion masses. For the one loop analysis (upper row) we used the LECs in \cite{Hanhart:2008mx} and for the two-loop analysis (lower row) we used the two sets A and D in \cite{Pelaez:2010fj}. We found that the dependence on the pion mass is again quite soft, specially for the $I$=2, $J$=0 channel and somewhat stronger at two loops than at one loop.

\begin{figure}
\begin{center}
\begin{tabular}{c}
  \includegraphics[scale=1.6]{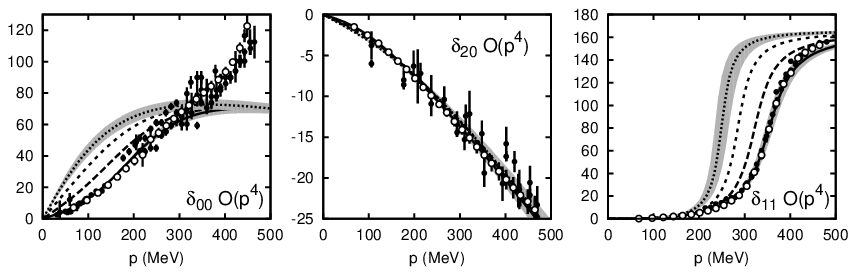}\\
  \includegraphics[scale=1.6]{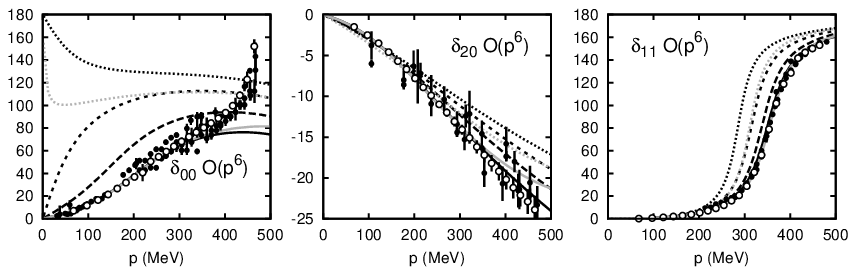}
  \end{tabular}
    \caption{$\pi \pi$ phase shifts from unitarized ChPT up to one loop (upper row) and to two loops (lower row). Different lines stand for different pion masses: continuous, long dashed, short dashed and dotted for $M_\pi=139.57,\, 230,\, 300$ and 350 MeV respectively. At one loop we show error bands only for the lightest and heaviest masses. At two loops no error analysis of the LECs has been carried out yet, thus we show results for the set of LECs called A in\cite{Pelaez:2010fj} and we add the results obtained with set D for the lightest and heaviest masses in light gray in order to give an idea of the size of the errors. Experimental data (rounded points) come from~\cite{experimentaldata} (black circles) and the precise model independent dispersive data analysis from \cite{GarciaMartin:2011cn} (white circles).}
  \label{fig:Uphaseshifts}
  \end{center}
\end{figure}

\begin{figure}
\begin{center}
  \includegraphics[scale=1.38]{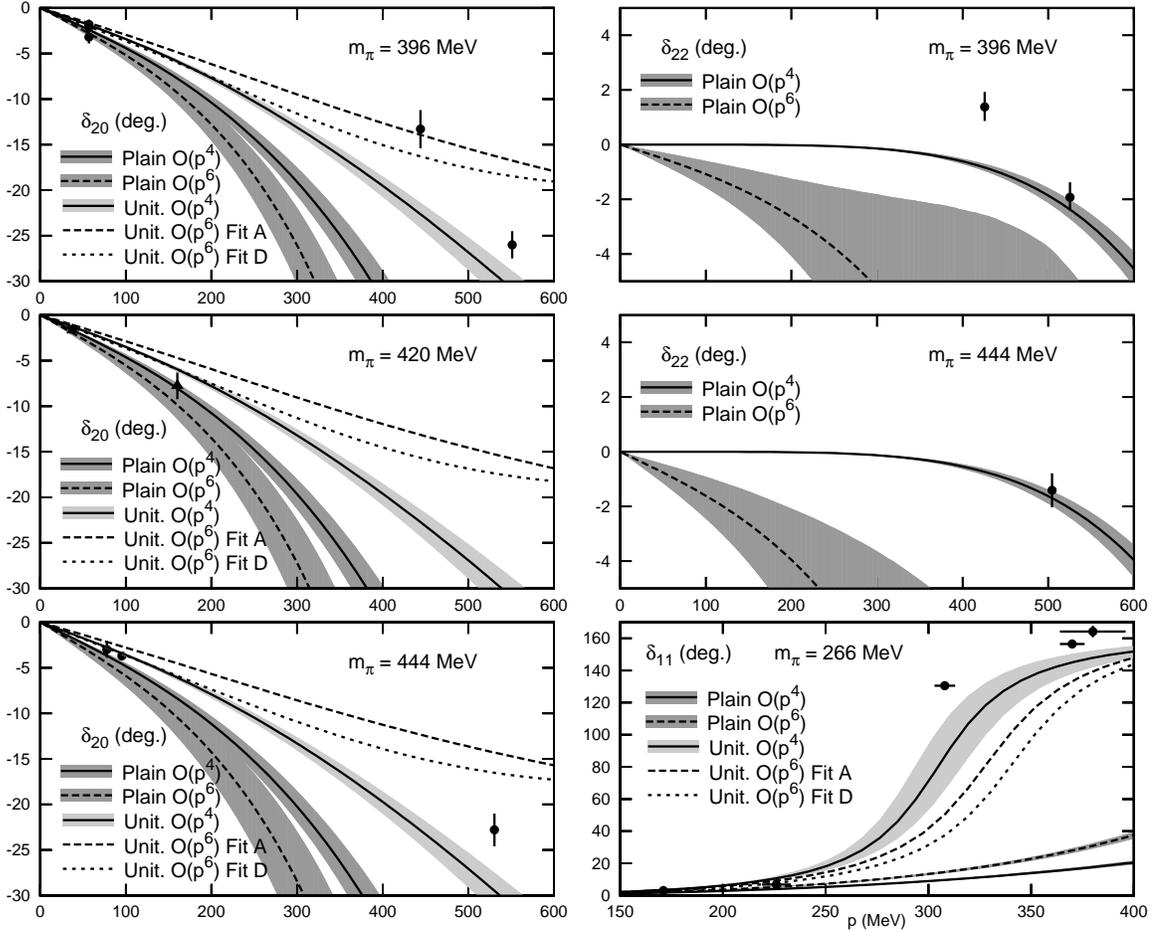} 
    \caption{\textbf{Left column:} One and two-loop phase shifts from plain and unitarized ChPT for the $I$=2, $J$=0 channel compared to 
lattice results coming from~\cite{Dudek:2010ew} (circles) and~\cite{Sasaki:2008sv} (triangles). \textbf{Right column, two upper panels:} One and two-loop phase shifts from plain ChPT for the $I$=2, $J$=2 channel compared to 
lattice results coming from ~\cite{Dudek:2010ew} (circles). \textbf{Right column, lower panel:} One and two-loop phase shifts from plain and unitarized ChPT for the $I$=1, $J$=1 channel compared to 
lattice results coming from ~\cite{Lang:2011mn} (circles).}
  \label{fig:lattice}
\end{center}
\end{figure}

\paragraph{Comparison to lattice results} In order to compare with lattice results~\cite{Sasaki:2008sv,Dudek:2010ew,Lang:2011mn}, we show in Fig.~\ref{fig:lattice} the $\pi\pi$ phase shifts in channels ($I$,$J$) = (2,0), (2,2) and (1,1) for higher pion masses, up to 444 MeV. Let us remark that this energy region is somewhat above the applicability limits of our method, which may provide precise results up to pion masses of, at most, 300-350 MeV. Therefore, our results at higher masses should be considered just qualitatively and, in fact, above $M_\pi=450$ MeV we do not even show them here (see~\cite{Nebreda:2011di}). For these proceedings we have added a new calculation of the $I$=1, $J$=1 phase shift at $M_\pi=266$ MeV in order to compare with the recent lattice results in \cite{Lang:2011mn} also presented at this conference. Standard ChPT shows a good agreement with lattice results below $p\simeq200$ MeV up to pion masses of 400-450 MeV, while a nice improvement above 200 MeV is found when using unitarized ChPT for the scalar and vector channels.

\vspace{-0.3cm}

\acknowledgements{%
We thank J. Dudek, D. Mohler and S. Prelovsek for lattice results and detailed explanations. Work partially supported by Spanish MICINN: FPA2007-29115-E,
FPA2008-00592 and FIS2006-03438,
U.Complutense/Banco Santander grant PR34/07-15875-BSCH and
UCM-BSCH GR58/08 910309 and EU-Research Infrastructure
Integrating Activity
``Study of Strongly Interacting Matter''
(HadronPhysics2, Grant 227431)
under the EU Seventh Framework Programme.
}

\vspace{-0.5cm}


%

}  


\end{document}